\documentclass[acmsmall]{acmart}

\usepackage{geometry}
\usepackage{longtable}
\usepackage{graphicx}
\usepackage{array}
\usepackage{multirow}
\AtBeginDocument{%
  \providecommand\BibTeX{{%
    \normalfont B\kern-0.5em{\scshape i\kern-0.25em b}\kern-0.8em\TeX}}}

\setcopyright{acmcopyright}
\copyrightyear{2025}
\acmYear{2025}
\acmDOI{XXXXXXX.XXXXXXX}

\acmConference[CSCW '25]{Proc. ACM Hum.-Comput. Interact., CSCW}{Oct 18-22}{Bergen, Norway}
\begin{document}

\title[Algorithmic Conspirituality]{\textit{"Whoever needs to see it, will see it":} Motivations and Labor of Creating Algorithmic Conspirituality Content on TikTok}

\author{Ankolika De}
\affiliation{%
  \institution{Pennsylvania State University}
  \country{USA}
}

\author{Kelley Cotter}
\authornote{Both authors contributed equally to this research.}
\affiliation{%
\institution{Pennsylvania State University}
\country{USA}
}
\author{Shaheen Kanthawala}
\authornotemark[1]
\affiliation{%
\institution{University of Alabama}
  \country{USA}
}

\author{Haley McAtee}
\affiliation{%
  \institution{University of Alabama}
  \country{USA}
}

\author{Amy Ritchart}
\affiliation{%
  \institution{University of Alabama}
  \country{USA}
}

\author{Gahana Kadur}
\affiliation{%
  \institution{Pennsylvania State University}
  \country{USA}
}

\renewcommand{\shortauthors}{De}

\begin{abstract}
Recent studies show that users often interpret social media algorithms as mystical or spiritual because of their unpredictability. This invites new questions about how such perceptions affect the content that creators create and the communities they form online. In this study, 14 creators of \textit{algorithmic conspirituality} content on TikTok were interviewed to explore their interpretations and creation processes influenced by the platform's For You Page algorithm. We illustrate how creators' beliefs interact with TikTok's algorithmic mediation to reinforce and shape their spiritual or relational themes. Furthermore, we show how algorithmic conspirituality content impacts viewers, highlighting its role in generating significant emotional and affective labor for creators, stemming from complex relational dynamics inherent in this content creation. We discuss implications for design to support creators aimed at recognizing the unexpected spiritual and religious experiences algorithms prompt, as well as supporting creators in effectively managing these challenges.
\end{abstract}




\maketitle

\section{Introduction}

%\begin{quote}
    %\textit{“Every time I think of the TikTok algorithm, I think toxic. I'll play nice, but this is toxic.” }
%-P25
%\end{quote}

The centrality of algorithms in orchestrating activity on social media and their perceived opacity \cite{10.1145/3637342, 10.1145/3290605.3300724} has invited a wealth of Computer-Supported Cooperative Work (CSCW) research on algorithmic folk theories (for eg., \cite{devitoRIPtwitter, eslamisocialfeeds}). Past research has also connected algorithms' obscured inner workings and unpredictability to conceptualizations of them as \textit{"magical"} \cite{10.1145/3613904.3641954, IJoC19289, singler2020blessed, doi:10.1177/2053951717751552, 10.1145/3544548.3581257}. Particular attention has been paid to TikTok, given its sudden, intense popularity, and its unique wholly algorithmically- driven content display interface — the For You Page (FYP) \cite{IJoC19289, 10.1145/3544548.3580649}. The prevalence of the idea among TikTok’s users that the FYP algorithm is both uniquely personal and perceptive led Cotter et al. \cite{IJoC19289} to conceptualize the phenomenon of \textit{algorithmic conspirituality}. Algorithmic conspirituality describes how TikTok users ascribe divine significance to especially well-tailored content shown to them in their FYP, even when they understand how the algorithm works \cite{doi:10.1177/13548565241258949}. This explicitly positions a video's message as sent to the individual viewer by a divine power, for example via phrases like \textit{"If you're seeing this, it's meant for you"} \cite{doi:10.1177/14614448231217425} and has been noted as increasing the persuasive appeal of videos\cite{doi:10.1177/14614448231217425}. 
 
Meaning-making through non-traditional epistemologies has been underexplored in Human-Computer Interaction (HCI) and CSCW literature \cite{10.1145/3411763.3450383}. Additionally, there is a scarcity of research on technology-mediated religious and spiritual spaces and practices \cite{10.1145/3613905.3651058, 10.1145/3290607.3310426}, particularly in the contexts of support seeking and relationship building \cite{hoefer2022faith}, which is surprising given the rapid growth of the new age spiritual movement \cite{IJoC19289}, alongside manifestations of digital spirituality \cite{partridge2021when}. 

We investigate how \textit{creators} of algorithmic conspirituality content comprehend and present their material as a unique form of expression. By examining their methods and beliefs we elucidate new perspectives on how algorithmically mediated communication intertwines with non-rational ideas about algorithms\cite{10.1145/3613904.3641954}. Informed by literature on labor \cite{10.1145/3544548.3580649}, and the challenges of online content creation \cite{aldous2019challenges}, we further elucidate the implications of content creation in relation to unique interpretations of algorithms \cite{10.1145/3476046, 10.1145/3359321}, particularly in how creators are affected by them. We ask:

\textbf{RQ1: How do creators understand, interpret, and perform algorithmic conspirituality and what expectations do they have in producing such videos?}

\textbf{RQ2: How are creators affected by creating algorithmic conspirituality content?}

Our findings show that creators produce algorithmic conspirituality content driven by spiritual, relational, and strategic motivations. They emphasized \textit{intuitive spontaneity}, reflecting chance-based views on the algorithm’s workings, which shaped their content. These practices aimed to connect with others and offer \textit{support}, using the algorithm to reach the right audience. However, creators also noted that the nature of their content often triggered overwhelming and intrusive interactions, fueled by genre expectations, resulting in considerable emotional and affective labor.

Overall, we contribute to HCI and CSCW research in three major ways. First, we reveal how the interplay between creators' beliefs and TikTok's algorithmic environment enhances the dissemination and reception of spiritually and religiously infused content for users. Next, we highlight how creators' strategies for personalized communication (\textit{"you-centric"} messaging) on TikTok leverage algorithmic mediation to produce a sense of open-ended context. Finally, by examining the relational nature of persuasive content and its amplification of creators' responsibilities, we complicate the affective labor they engage in and contribute to CSCW and HCI design by proposing features that help creators manage this labor.

\section{Background}
% algorithmic consp, visibility pursuits
% Here, we focus on the mystical and conspiratorial narratives around AI that affect user understanding. These narratives and folk theories influence content creators' self-presentation, audience relationships, and multifaceted labor within algorithmically mediated environments.

\subsection{Algorithmic Conspirituality and Human-Computer-Divine Interaction}
\label{algohci}

The ubiquity of artificial intelligence (AI)-driven applications has created a growing disconnect between the complexity of these technologies and their comprehensibility among general users. The perceived powers enabling AI have often been examined through lenses of spirituality, transcendence, faith, and religion. Here, \textit{spirituality} refers to \textit{"a sense of deep connection to something larger than oneself, whatever that may be"}, encompassing not only ideas beyond religion but also more personal, overarching beliefs in something beyond the self \cite{10.1145/2468356.2468754}. This spirituality differs from religion, which has been defined as more organized and structured around common beliefs shared by groups \cite{10.1145/3371382.3380736}. Other works have built on this distinction to explore \textit{transcendence} which draws on common spiritual concepts found in Western philosophies, where “\textit{meaning and communion surpass the ego and personal boundaries}” \cite{10.1145/3027063.3048434}, rather than adhering to a specific religious doctrine. Prior work has also shown that spirituality can manifest in interactions with spiritual apps, such as tarot reading and wellness or mindfulness systems \cite{rosner2022spirituality, 10.1145/3643834.3660735}. 

An overt example of the content we are interested in is the \textit{"if you see this"} genre of videos on TikTok \cite{IJoC16414}, which draws from contemporary interest in new-age spiritual practices, combined with the prevalence of conspiratorial thinking. In fact, prior work has defined- \textit{conspirituality} as a philosophy wherein - \textit{"nothing happens by accident, nothing is as it seems, and everything is connected"} \cite{ward2011emergence}. Based on this idea, Cotter et al., \cite{IJoC19289} introduced \textit{algorithmic conspirituality}- "\textit{a spiritualizing beliefs about algorithms, which emerge from occasions when people find personal, often revelatory connections to content algorithmically recommended to them}" \cite{IJoC19289}. This scholarship investigated TikTok videos that introduced a \textit{meant to be} discourse - with captions or content like: \textit{"If you’re seeing this, it’s because you set this reminder in place. You asked me to bring you this reminder when you need it..."} \cite{IJoC19289}. Kanthawala et al. \cite{doi:10.1177/14614448231217425} explicated algorithmic conspirituality's persuasive capacities in altering viewers' beliefs and behaviors. Algorithmic conspirituality content also benefits from persuasion and promotion, as seen in mass media and advertising research. The parasocial relationships \cite{horton1956mass} this content facilitates between the creator and viewer and viewers' perception of the algorithm knowing them intimately \cite{IJoC19289, doi:10.1177/20563051221144322} likely increase the perceived authority of messages \cite{doi:10.1177/14614448231217425} as both the creator and the algorithm act as gatekeepers \cite{soffer2021algorithmic}. 

Additionally, alternative modes of sensemaking of algorithms in this context have identified mystical elements, also defined as the "\textit{possibility of acquiring ineffable knowledge attributed to the divine}" \cite{10.1145/3563657.3595990}. In particular, Cotter et al.'s \cite{doi:10.1177/13548565241258949} work found that TikTok users expressed \textit{reflexive ambivalence}, recognizing both mystical and rational aspects in algorithmic conspirituality videos.

In a tangential line of work, Hoefer et al., \cite{hoefer2022faith} coined \textit{faith informatics} to study "\textit{systems that facilitate growth in individuals’ systems of meaning-making}". Faith is a broad term that is often used interchangeably with \textit{trust}. While epistemologically rooted in various disciplines, we draw from \cite{bishop2010faith} to explicate faith as trust in ideas and events that are not easily explained by other means. Likewise, faith informatics \cite{hoefer2022faith} has been encouraged to successfully facilitate social interactions \cite{fowler1981stages}. Extending this work, we hope to see how the creation of algorithmic conspirituality content both supports this goal and goes beyond it to foster connections.

When examining content creation within AI-mediated spaces, prior work has established the importance of mediating support through spirituality \cite{10.1145/3290607.3310426}, faith \cite{hoefer2022faith} and religious experiences \cite{10.1145/3491101.3519856, 10.1145/3563657.3596029, evolvi2022religion} with AI-driven tools \cite{10.1145/3449117, 10.1145/3656156.3663723, 10.1145/3555162}. However, these discussions \cite{10.1145/3613905.3651058} within HCI remain scarce \cite{10.1145/3613905.3651058}.

While existing research has explored the mystical and opaque framing of AI \cite{Campolo2020Enchanted}, algorithms \cite{IJoC19289}, and big data \cite{doi:10.1177/14614448241230923}, including concepts such as algorithmic conspirituality, there has been limited investigation into how content creators themselves understand and engage with these frames. We extend this research within the HCI and CSCW literature by examining how creators' understanding of algorithms and algorithmic conspirituality influence their content creation strategies and shape their relationships with audiences. 

\subsection{Understanding Algorithms, Audience Relationships, and Self Presentation}

Folk theories describe popular, intuitive \textit{“theories”} about how something--technology, in this context--works, as formulated by everyday people \cite{eslamisocialfeeds, 10.1145/3170427.3186320}, rather than experts \cite{doi:10.1177/0163443720972314, doi:10.1177/2053951720923377}. In the absence of understanding about the functionality of algorithms on these sites, folk theories essentially \textit{"fill in the gaps"} in everyday users' knowledge of the algorithms as they make sense of their experiences within these platforms \cite{devitoRIPtwitter,10.1145/3173574.3173694}. 

Content creators’ folk theories  influence their content creation, including the \textit{“modes of self-presentation, tone of voice, choice of content covered, words and sentence structures used” or otherwise} \cite{bishop_anxiety_2018}. For instance, researchers explained how TikTok's algorithmic experiences helped form tight-knit communities among marginalized individuals, shaping their personal identities \cite{simpson_for_2021}. Related work has found that algorithmic folk theories particularly shape other communities' behaviors as well, including groups of Black TikTok creators, women, and LGBTQ+ groups \cite{10.1145/3610169, DeVito_algotrap, 10.1145/3544548.3580970}. Algorithmic curation and folk theories also impact individual's identity work and presentation of the self as they mediate their relationships with their audiences \cite{10.1145/2998181.2998192, 10.1145/3173574.3173694, devito2020presenting}.

Content creators, regardless of self-identifying as social media "influencers," use similar audience management practices \cite{marwick_i_2011} to conjure an imagined audience. Previous research has examined how individuals conceptualize their audience, either as a monolith or in more personalized, niche terms \cite{litt_imagined_2016}. Social media users often reported not considering their audience’s characteristics or thinking specifically about who would view their posts \cite{litt_imagined_2016, doi:10.1177/20563051211035692}. Niche imagined audiences are typically conceptually homogeneous, such as groups united by common interests like sports fandom \cite{litt_imagined_2016, doi:10.1177/20594364231222976}.

Content creators also consider the types of audiences they are building based on the social media platform  \cite{hansen2021digital, gardner2016new}. Additionally, they prioritize audience management across multiple platforms \cite{ma_multi-platform_2023} to cultivate fandom as a way to build their brand and a revenue stream. Other research has characterized audiences in terms of \emph{publics} with relationships connected to self-identity and collective belonging \cite{lindtner_towards_2011}. Our study examines how creators interpret algorithms, engage audiences, and build relationships within algorithmically mediated content economies.

\subsection{Managing Labor as Content Creators}

HCI and CSCW scholarship has examined content creators' labor in navigating algorithmically driven spaces \cite{10.1145/3462204.3481784, 10.1145/3544548.3580649}. Specifically, these studies focus on understanding how content creators interact with algorithms \cite{10.1145/3359321, 10.1145/3173574.3173694}, and how the process of algorithmic mediation impacts them by often demanding additional labor \cite{10.1145/3613904.3642173, 10.1145/3544548.3580649}. Here, we explicate the different kinds of labor that creators navigate to thrive within algorithmically mediated environments.

Creators employ various strategies, such as adhering to trends \cite{10.1145/3613904.3642173, doi:10.1177/20563051211021368}, engaging in advertisements \cite{10.1145/3613904.3642173, doi:10.1177/14614448221081802}, and maintaining constancy \cite{haug2024content} among others, to remain visible and ensure that their content is algorithmically recommended to users (Also see for eg., \cite{silberman2023content, 10.1145/3637385, 10.1145/3472714.3475826, doi:10.1177/1461444814541527}).  Additionally, creators must manage their personal well-being while navigating platforms and engaging in interactions, which leads to significant amounts of emotional labor \cite{hochschild2012managed, doi:10.1177/20594364221096498, 10.1145/3290605.3300390, 10.1145/3544548.3581386}. Scholars have also found emotional labor to often be both impacted by and lead to the commodification of authenticity \cite{illouz2019emotions}- as \textit{emotion}, is \textit{"subject to acts of management"} \cite{hochschild1979emotion}. Moreover, creators also manage the repercussions of their content, including user responses \cite{10.1145/3613904.3642148, haug2024content}, and in managing content amidst platforms' moderation decisions \cite{10.1145/3500868.3559445}- engaging in affective labor \cite{doi:10.1177/20594364221096498, hochschild2012managed, articleosaka}. Hardt \cite{hardt1999affective} posits that affective labor underpins \textit{"immaterial labor"}, where activities such as "caring" yield products perceived as devoid of tangible worth in the market economy. Thus, while creators' do not gain any direct material benefit from their affective labor, they must perform this work to build an audience.

Altogether, this creative labor entailed by content creation for algorithmically-mediated environments demands a business-oriented mindset to remain competitive \cite{doi:10.1177/2056305119879672, 10.1145/3492841}. Put differently, being a content creator encompasses a broader range of labor practices than merely creating content. Indeed, as scholars argued, \textit{being creative} is a process inherently fraught with tensions arising from platform constraints related to logistics and visibility, among other factors \cite{10.1177/14614448211027961}. Our study expands upon existing research on the challenges creators face in navigating labor dynamics while innovating within distinct genres, influenced by unconventional interpretations of algorithms.

%\subsection{Audiences, Relationships and Self Presentation}

%Developing a working concept of who will be receiving, viewing and potentially following or sharing the content is a primary element in establishing the relationship.

\section{Methods}

We conducted semi-structured interviews with 14 creators of  \textit{"if you see this, it is meant for you"} format algorithmic conspirituality \cite{IJoC19289} videos. Simultaneously, we conducted qualitative iterative-inductive-thematic analysis \cite{braun2012thematic} to report our findings.

\begin{table*}[t]
\centering
\resizebox{\textwidth}{!}{%
\begin{tabular}{ccp{0.9\textwidth}} % Adjust the column width further
\hline
Participant ID & Video Type & Summary \\ \hline
1 & Religious & \small The creator states that they prayed before recording the video asking to be put on the fyp of someone really beautiful, which is the viewer. They remind the viewer that Jesus loves them and encourage them to read scriptures. \\
2 & Spiritual & \small The creator states that if the viewer is seeing the video on a specific date, then something they've been wanting to happen for a long time will finally happen. They state the viewer is worthy of this, and say to save the video to claim the message. \\
3 & Spiritual & \small The creator states they have a message the viewer needs to hear and instructs them to put their phone to their ear. The creator then states they see the viewer coming out of the rough period they’re in, and encourages them by saying they are a lot stronger than they think. \\
4 & Religious & \small The creator states that they didn’t put any hashtags so if the video comes across someone’s page, they feel like God wants them to pass a message to the viewer. They then say that the viewer’s presence is enough, and that they themselves are enough. They say this is the reason people want to talk to them, and that they bring light to people’s day because of God’s presence in them. They say the viewer’s purpose is their presence and say they have to take care of themselves. They say to be grateful to situations God has put them in. \\
5 & Spiritual & \small The video features the creator dancing in the background and text on screen saying they are manifesting that the money both them and the viewer spend comes back to them. The video features the numbers 444 and 888 and also repeats the sentence three times with “x3.” \\
6 & Secular & \small The video features the creator telling the viewer that this is a sign to take Instagram seriously, and that anyone can do it. Then they show before and after pictures of their own Instagram account, as well as stating that they got paid for it as well. \\
7 & Secular & \small The video features the creator lifting weights, and the text on screen tells ladies that this is their sign to build their back if they want an hourglass figure. \\
8 & Secular & \small The video features text at the beginning saying the video is a sign to visit Mackinac Island in Michigan. They then show various features of the Island people can come see including the downtown, turquoise waters, biking, horse-drawn carriages, and historical sites like homes, hotels, and forts. They say the viewer can come by ferry or plane, and write “see you there” at the end. \\
9 & Spiritual & \small In this video, the creator tells the viewer to look around and connect with their senses to notice that they have everything they need and desire. Even if the viewer wants to rebut the statement, the creator states that in this moment, they are breathing and their heart is beating and they are consciously listening to the video, thus they have everything they need. They state that once this is recognized, all their dreams and desires will become reality. \\
10 & Religious & \small The person in the video takes a rubber band and stretches it and asks if anyone has felt like the rubber band—stretched and at their breaking point. She says this is because in order for the viewer to get where they need to go, God had to stretch them. She states they were stretched for a purpose, and they won't break. \\
11 & Spiritual & \small The creator says that the viewer needs to write down these two dates because really significant change is coming to their life on July 28th or September 16th. They say these two dates keep coming to them so they decided to make a TikTok about it. They say the dates will be significant in their soul’s path and will change your destiny. \\
12 & Spiritual & \small The video shows someone lighting a candle, and various dollar amounts appearing on different parts of the video. The audio states that a spirit told them to tell the viewer that if they use this sound they will receive a huge windfall, and to not delay. \\
13 & Spiritual & \small This creator holds up the “come together” tarot card, and on the screen has written that this video is meant for you. They tell the viewer congratulations because the person the viewer is manifesting, possibly an ex, will get back together with them, and they will probably get a happily ever after ending. \\
14 & Spiritual & \small The creator uses the text, "If you are watching this right now, you are not alone. I’m here, and I'll always be here-- while nodding their head in agreement in the video, and making eye contact with the viewer-- essentially narrating that the viewer should not feel alone. \\
\hline
\end{tabular}%
}
\caption{\label{tab:vidsum} Video Summaries and Types: This table provides a description of the videos created by the interviewers in our study and categorizes these videos according to their content.}
\end{table*}

\subsection{Data Collection}
%Data Collection
    % Collecting emails
    \subsubsection{Participants and Recruitment}
Interview participants were mainly recruited via purposeful sampling \cite{suri2011purposeful} with the TikTok "search" function. To identify the most relevant creators for participation, the authors conducted a targeted search for content within the sphere of algorithmic conspirituality. Search terms included the phrases: \textit{"This is a sign"},\textit{"If you see this"}, and \textit{"Meant for you"}. These search terms indicate the consistency of returned videos with algorithmic conspirituality, as these phrases communicate the creator's implied belief that their videos will reach a specific imagined audience by way of the FYP algorithm \cite{doi:10.1177/14614448231217425}. Table \ref{tab:vidsum} describes the videos' content of the creators we interviewed, while Table \ref{tab:identities} summarizes the religious and spiritual identities of the participants, as they could be gleaned from their discussions in their interview. All participants provided consent per institutional IRB guidelines and received \$35 USD Visa or Amazon gift cards. The demographic details of participants who completed a follow-up questionnaire are presented in Table \ref{tab:demographics}.

\begin{table*}[t]
\centering
\begin{tabular}{cc} 
\hline
\textbf{Participant ID} & \textbf{Personal Beliefs} \\ \hline
1 & Religious \\
2 & Unclear \\
3 & Spiritual \\
4 & Religious \\
5 & Spiritual \\
6 & Spiritual \\
7 & Unclear \\
8 & Unclear \\
9 & Spiritual \\
10 & Religious \\
11 & Spiritual \\
12 & Spiritual \\
13 & Spiritual \\
14 & Unclear \\
\hline
\end{tabular}
\caption{\label{tab:identities} Video Creators' Personal Beliefs: This table summarizes the religious or spiritual identities of the participants, based on their interviews. The rationale behind categorizing personal beliefs is based on the explicitness and nature of the beliefs expressed by creators. Religious beliefs were categorized when creators openly identified with specific faiths or practices, as seen in the following example: \textit{"So my TikTok account is known as a faith-based Christian content account"} (P1). Spiritual beliefs were categorized when creators mentioned beliefs in forces or energies beyond the material world, without aligning with a specific religions, as in \textit{“I don’t want to get too deep. But my twin brother passed away three years ago. And whenever I do a message like that, I feel like it’s him talking to me.”} (P3). Content without clear religious or spiritual context was categorized as unclear. Here, no explicit mention of beliefs manifested in the interviews.}
\end{table*}

\begin{table*}[t]
\centering
\resizebox{150mm}{!}{%
\begin{tabular}{ccccccc}
\hline
ID & Age & Gender & Race/Ethnicity & Education & Gross Personal Income & No. of Followers \\ \hline
1 & 23 & Male & White & College diploma & \$10,000 up to \$19,999 & 330,000\\
2 & 28 & Male & White \& Black or African American & College diploma & \$50,000 up to \$59,999 & 157,000\\
3 & N/A & N/A & N/A & N/A & N/A & N/A\\
4 & 23 & Female & Black or African American & College diploma & \$10,000 up to \$19,999 & 116,700\\
5 & 25 & Female & White \& Black or African American & Associate's degree & Up to \$10,000 & 8,000 \\
6 & 19 & Female & Hispanic, Latino, or Spanish & Some college & \$30,000 up to \$39,999 & 1,000,000\\
7 & 21 & Female & Asian & College diploma & More than \$150,000 & 26,000\\
8 & 34 & Female & White & Master's degree & \$100,000 up to \$149,999 & 205,000 \\
9 & 30 & Female & Asian & Master's degree & \$90,000 up to \$99,999 & 110,900 \\
10 & N/A & N/A & N/A & N/A & N/A & N/A\\
11 & 52 & Female & White & College diploma & \$70,000 up to \$79,999 & 129,000\\
12 & 27 & Female & White & College diploma & \$30,000 up to \$39,999 & 16,400\\
13 & 24 & Female & White & College diploma & Up to \$10,000 & 330,000\\
14 & N/A & N/A & N/A & N/A & N/A & N/A\\
\hline
\end{tabular}%
}
\caption{\label{tab:demographics}Participant demographics}
\end{table*}
    % Reaching out

    %interviews

    \subsubsection{Interviews}
Our interview protocol, started with getting to know the creators by asking, "\textit{What type of content do you post?}" and "\textit{What “sides” of TikTok are you on?}". We then presented participants with the specific video that led us to contact them, as it formed the basis for our questions about their decision to create this kind of content. Sample questions included, "\textit{Who is the 'you' that you are addressing?}", "\textit{What are you hoping to achieve with this format?}", and "\textit{What kind of audience are you trying to reach?}" Next, we explored their video creation process to understand their strategies and motivations. Example questions included, "\textit{How do you go about creating your videos—both this kind and other formats?}" and "\textit{How do you think the algorithm helps in spreading these videos?}" Finally, we addressed questions about the labor involved in maintaining their role as creators, including their perceived responsibility to their audience. Example questions included, "\textit{What kind of responses do you get for your video?}" and "\textit{How do you interact (or not) with your audiences?}."

%Our protocol evolved in response to the data, guiding our exploration into new themes.

%Data Analysis
    %Building the code book
        %Rounds of open coding and Deliberation
        %Finalized the codebook
    %Doing the analysis
\subsection{Data Analysis}

After data collection, all interviews were transcribed and anonymized. A categorization of the videos that were used to contact creators is given in table \ref{tab:vidsum}. Religious videos were identified by clear references to organized belief systems, such as mentions of specific religious figures (e.g., "Jesus") or practices (e.g., prayer) \cite{10.1145/3613905.3651058}. When a creator referenced "God" without additional context, we determined it to be tied to religion. Our categorization of spiritual videos emphasized personal beliefs like manifestation or self-empowerment without citing a religious tradition, while secular videos used the algorithmic conspirituality format without explicit religious or spiritual content \cite{rosner2022spirituality, 10.1145/3643834.3660735}. It is essential to clarify that Table \ref{tab:vidsum} represents the content of the videos, while the beliefs of the creators were assessed based on interviews and are in table \ref{tab:identities}, contingent upon whether participants organically introduced their religious or spiritual beliefs. Notably, the study did not initially aim to focus heavily on the religious or spiritual beliefs of creators. However, it became evident that many creators in the algorithmic conspirituality genre produced videos of this nature and also held personal beliefs related to it.

The interview transcripts were analyzed by conducting an iterative, inductive thematic analysis \cite{braun2012thematic, Braun2019}. This approach involved generating themes directly from the data (inductive) \cite{doi:10.1177/109821409101200108}, repeatedly refining themes to clarify them and achieve team consensus (iterative) \cite{doi:10.1177/1609406917733847}, and developing a codebook with focused themes \cite{saldana2013coding} that guided our findings. An example of the iterative process was the refinement of descriptive level codes such as \textit{caring for audiences} and \textit{responding to viewer needs} folding into affective labor as a focused code. Each co-author reviewed 1-3 transcripts in an initial round of open coding. Discussions of these codes led to the development of preliminary descriptive themes. Following this, a second round of coding took place, where each team member coded 1-3 new-to-them transcripts that they had not previously reviewed. This second round of coding utilized the preliminary descriptive themes generated in the earlier round. The research team met periodically to deliberate and clarify any disagreements. This collaborative effort resulted in the final codebook, which was subsequently used to code all the interviews. The final round of coding was distributed among all members of the coding team.

%All team members participated in the coding process and contributed to the discussion of the generated themes.

\section{Findings}
Our analysis elucidated two overarching themes: displaying and performing algorithmic conspirituality (RQ1), and the impact on creators of doing so (RQ2). Each theme is broken into sub-themes and elaborated in more detail below. A summary of this section can be found in table \ref{tab:findingsum}.

\subsection{Displaying and Performing Algorithmic Conspirituality}

Content developed within this genre primarily engaged with religious and spiritual themes. Creators either emphasized their own beliefs in framing how and why \textit{this} content was reaching their audiences, or engaged in audiences' identities. They reflected on how their beliefs shaped their views on the algorithm and audience engagement. Creators who felt their content was spiritually or religiously driven also shared how they relied on their deductions about how the algorithm was working at a particular time to create content that would reach their audience. In doing so, some creators used complex strategies to suit the algorithm and engage viewers, while others described a kind of \textit{non-strategy strategy} premised on faith in algorithmic recommendation. 

\subsubsection{A Divine Intervention}
\label{sec:divine}

\begin{quote}
    \textit{I'm so fired up from the prayer before with the Lord, I really feel his presence. And then that three, two, one, go. And I just kind of say what comes to mind, which comes out of my mouth which I've learned is just allowing, I believe the Holy Spirit to speak through me and to people across the screen (P1).}
\end{quote}

Like P1, some creators who incorporated elements of algorithmic conspirituality in their videos attributed their viral success to a \textit{divine intervention}-- as an ongoing, dynamic interaction between a \textit{"higher power and themselves"}, believing that their actions could subtly influence the divine, resulting in miraculous acts woven into everyday life \cite{10.1145/3563657.3595990}. Here, "divine" does not necessarily refer to a monotheistic belief but rather to something beyond the human realm—an otherworldly presence that embodies faith, spirituality, and religion, which may take different forms for different creators.  Participants also believed that a higher power guided their content, a notion supported by their acknowledgment of viewers who they perceived as personally touched by their messages. Such creators expressed three motivations for producing algorithmic conspirituality content: \textit{serving as messengers of a higher power}, \textit{fulfilling their own spiritual and religious needs}, and \textit{testing their relationship with the higher power}.

First, for some creators, producing algorithmic conspirituality videos was tied to their belief in being a messenger of something \textit{beyond themselves}. Some among them were religious in that they developed \textit{personae} similar to that of \textit{pastors}-- "\textit{using media as it is designed, with careful role performance and attention to the audience}", to share what they perceived as messages from God \cite{checketts2018persona}. These messages were often unplanned and spontaneous, with the content and timing inspired by what they felt as divine guidance. When asked about their reason for using the specific format P4 noted, \textit{"In that moment that day, I’m like, whoa, that was nothing but God"}, providing some background for this belief by sharing- \textit{"I had been getting closer in my relationship with God as well. And I’ve been kind of on this journey of just trying to allow God to just work through me"} (P4). 

At times, interviewees also saw the specific content of their videos as a result of their own religious role, in documenting their own religious experiences and sharing it with others. For example, P1 cited his \textit{"sole goal"} as \textit{"shar[ing] the gospel"} . P4 also contributed to this, 

\begin{quote}
    \textit{And so before I made that video, I had prayed and I was like, OK, God, just work through me to bless whoever needs to be blessed within this video that I’m making. Work through me to be able to just sing in a way that I may have never been able to sing before.} 
\end{quote}

P3 shared a different experience, saying, 

\begin{quote}
    \textit{[M]y twin brother passed away three years ago. And whenever I do a message like that, I feel like it's him talking to me. That's where I dive into, and I see it in my head [..] If I don't get the goosebumps when putting up a message, then it's not going to do well. I have to feel it deep inside.} 
\end{quote}

While not explicitly referencing God, P3 described a spiritual experience where he perceived his brother speaking through him. This suggests a sense of otherworldly influence and points to a personal experience that may align with religious beliefs in the afterlife, but may also reflect transcendence. 

Beyond serving others, these videos also supported the creators' own spiritual and religious well-being, often spilling over into other forms of manifestation and New Age beliefs. P4 shared, \textit{"It literally feels like I’m talking to myself because that’s what I needed to hear at that moment"}. Here, the creator tried to address her own needs and beliefs-- and in turn provide an emotional support through affirmations \cite{10.1145/3402855}. Additionally, another creator, P10 shared "\textit{[I am] just using things that I've learned from God that's helped me that might be able to help someone else}"- and through that she adds how it became a conversation that she wanted to pursue anyhow, and thus it did  not involve an active act of \textit{create [ing] content}. She continued, "\textit{So it's a thought that I'm going to have [...] or I'm going to have this conversation, or I'm going to read this nugget, and I'm going to just record it.}". This process of creating content served a dual purpose: it not only helped them communicate messages to others but also allowed for a fulfillment of their emotional needs as a result of their affirmations. 

%By articulating their thoughts and feelings through these videos, they \textit{affirmed} their own needs, effectively nurturing their personal spirituality while also engaging with a broader audience.

P4 also shared how she tested her relationship with God through her content, in understanding her connectedness with religion. For instance, she said,

\begin{quote}
    \textit{I really wanted to take a leap of faith with this and just really see like, ‘Okay, if you’re really going to work through me and if you’re really going to put this on somebody’s feed that needs this, I’m not going to include anything that has to do with the video, and whoever needs to see it will see it’.}
\end{quote}

Seeing God as working through the algorithm, P4 wanted to have \textit{faith} that it could interpret her intent and deliver content to recipients who "\textit{needed}" it: "\textit{And if it’s the algorithm that puts it on there, okay. If it’s God that specifically sends it to the people that need to see it, that’s even better}". P1 additionally explicated the perceived divine will of the algorithm- \textit{"I pray right before I play over the videos that the Lord only show this exact content to who is supposed to see it in Jesus’ name"}. Thus, creators embraced the use of religious, spiritual and more secular ideas to discuss their own shaping, producing and explication of algorithmic conspirituality content, much like the \textit{mystical and rational} elements that viewers ascribed to it \cite{doi:10.1177/13548565241258949}. 

%This underscored creators' openness to both algorithmic and spiritual explanations for how their messages might reach their target audience. 

Notably, other manifestations of spiritual motivations distinct from monotheistic religions were included, P5, a tarot card reader, who shared that she "\textit{operate[d] from energy}," explaining:

\begin{quote}
\textit{And then one day, I just had this weird feeling, and something told me to sit in front of my camera and start recording and post one of my first videos and I did. And it did okay, but by the third video, I was starting to gain traction, so that’s when I kind of knew that I was on the right path in doing what I felt I was supposed to be doing.}
\end{quote}

P5's  spiritual message capitalized on the algorithm and TikTok's response to validate it. Thus, while she believed she was a messenger of this higher power, she also felt validated by user responses, in addition to her own beliefs.

\subsubsection{Personalized Engagement}
\label{sec:peson}

Our interviewees reported that the \textit{"you-centric"} structure and personalization of algorithmic conspirituality videos increased their likelihood of audience engagement, providing an effective \textit{"hook"}, explaining that this video format enabled them to establish \textit{stronger connections} with viewers, as well as remain motivated to continue producing content. These \textit{strategic} and \textit{relational motivations} coexisted with the spiritual and religious motivations highlighted in section \ref{sec:divine}.

In the context of the algorithmic conspirituality format acting as a hook to amplify their content, P6 shared, \textit{"I feel like if you don’t get people’s attention within the [first] three, five seconds [...] you literally swipe."} P3 further noted- \textit{"When you say certain phrases in a video, the AI catches it on TikTok."} 

P10 similarly recognized the effectiveness of the format as a hook, and explained her use of it as inspired by the rhetorical technique of a \textit{pastor} she admired, 

\begin{quote}
    \textit{There was a pastor. [... when I] went to his conference. And I realized when he was speaking to over 10,000 people, but he was saying, "You." He wasn't saying, "Y'all." He wasn't speaking to the group. He was speaking to the person.}
\end{quote}

She further added, saying \textit{"y'all"}, \textit{"gives room for insecurity"}, and may make the audience think \textit{"'Oh, she's not talking to me [..] (if someone says) y'all, I don't think I'm doing good, so she's probably meaning somebody else.'"} It is important to note that, while P10 also produced content about Christianity and her own beliefs, when speaking about the concept of algorithmic conspirituality, she did not explicitly state religious motivations. However, she was influenced by her religious life, such as drawing inspiration from her pastor to strengthen relationships, which contributed to her pursuit of this relational motivation. P10 explained her content structure further: 

\begin{quote}
    \textit{So automatically, when [viewers] see that ‘this is for you,’ [they think] maybe it is, right? If they see that up there and it’s-- first of all, I’m talking to the person, not to people, right? So it’s like me and you are having a conversation. I’m calling you. I’m not just trying to put out a broad spectrum to everybody, so it makes it personal. And everybody likes things that are personal.}
\end{quote}

Elaborating, P10 continued that when someone was looking for specific content and encountered "you"-centric phrasing, they were more likely to be engaged. This personalization strengthened the \textit{hook}, making it more effective in capturing and retaining the viewer's attention.  

\begin{quote}
     \textit{So if I say [this is for you, etc.], [..]and it catches you, you’re not going to scroll past it if you’re looking for something that you know you want to be connected to because everybody is looking for a message for them, right, even if they don’t realize it.} 
\end{quote}

Another creator explained how his use of the algorithmic conspirituality format came to define his content, eventually becoming a requisite for content amplification and recognizability among his audiences: 

\begin{quote}
    \textit{It got to a point where if I didn’t say that [If you're seeing this video, it is meant for you], my video never performed as well. And the crazy part about all of this is that people grow accustomed to the hook that you have and what you say. }(P2) 
\end{quote}

However, he also acknowledged that initially using this format completely aligned with his pursuit of authenticity:

\begin{quote}
   \textit{I saw someone’s video where they were-- it was a different niche. She was saying the same hook of, 'If you’ve seen this today, this message is for you.' But she applied it towards a breakup and giving a message from there [..] I stayed to watch it. I’m not going through a breakup, but I just realized I watched this all the way through [..] So I created a message like that [..] And it just started going crazy. And so many people were liking it. They were commenting like, 'This is what I needed to see. Thank you for this.' [..] I was like, This is amazing on both ends. This is me authentically creating my message, doing what I want to do."}
\end{quote}

P4, who shared that she was highly motivated for religious reasons stated- \textit{"I would go with the algorithm just to kind of get on the wave of things and then draw certain people to my page and get my interactions and stuff up"}. This highlighted the importance of adapting her content to trends, noting, \textit{" You can always put your own spin on different things and be original with trends. But yeah, just doing certain trends because that’s what most people’s eyes are on"}- suggesting that there was a  blend of performativity alongside any purported authenticity regarding religious and spiritual motivations and beliefs.

Creators' strategic motivations also intersected with \textit{relational} ones, as they intentionally sought to connect with and positively impact their audience. Explaining this, P14 described his audience, saying

\begin{quote}
     \textit{I would say the audience that don't have nobody else to motivate them, to tell them that everything's going to be okay, just the audience to heal [..] that's going through a lot right now don't have nobody to motivate them, [and] feel like giving up on life.}
\end{quote}

Likewise, P3 emphasized a desire to impress a sense of belonging among viewers:

\begin{quote}
\textit{The fact that one person, one person hears that because this is for you - 'Me? Really?' And they'll say, 'I feel like you're in my head.' It’s offering somebody a safe place is very, very important to me.}  
\end{quote}

P10 similarly noted that offering support had always been the goal, \textit{I truly felt like when I talked to people [...] when I got on TikTok, my goal was [to] help one person",} highlighting how the format helped reach audiences, but that impacting even \textit{one person} was sufficient. P11, who was also spiritually motivated and whose niche was tarot readings, added that,

\begin{quote}
    \textit{The client base is more important to me than the views and the follows and all of that stuff. The one-on-one people that I get to meet, the people that I work with, the people that become coaching clients, the people that I can help support through a tough time. }
\end{quote}

Here, P11 positioned her content creation as primarily motivated by connections, while highlighting the importance of this for her success. 

The intertwining of \textit{strategic} and \textit{relational} motivations reflected a holistic approach to algorithmic conspirituality on TikTok, where creators leveraged platform dynamics to amplify their relational impacts. These insights illuminate the multifaceted motivations driving creators in navigating TikTok's ecosystem, highlighting their strategic adaptability and relational engagement.

\subsubsection{Intuitive Spontaneity}
\label{sec:intu}

Our participants often described their creation process as \textit{spontaneous} and \textit{instinctive} particularly in the context of the perceived unpredictability of TikTok's FYP algorithm. Rather than meticulous planning or strategic analysis, they relied on their intuition and on-the-spot inspiration. For example, P2 explained that he could only achieve actual visibility after \textit{feed[ing]} the algorithm, so that it accepts him and amplifies his work-

\begin{quote}
\textit{I had to kind of feed it [the algorithm]. In order for me to get to the door or even to be accepted, I had to make sure I was hitting all of those [unknown metrics]. And then it kind of was like, "All right, you're in the club."}
\end{quote}

While this might seem contradictory to the strategic engagement we describe above, we understand this form of \textit{intuitive spontaneity} as a \textit{non-strategy strategy} that creators felt worked for them. In other words, intuitive spontaneity represented a strategy characterized by an explicit absence of predetermined planning. As P3 explained,

\begin{quote}
    \textit{I know for a fact that [..] a lot of us are clueless. We don't know what is going on. We just try, we put it out there, and some catch, some don't. I'll have a video with 3 million views and then a video with 3,000.}
\end{quote}

This uncertainty surrounding TikTok’s algorithms reinforced the idea that any content could succeed, fostering a sense of serendipity that led creators to embrace unpredictability—ironically, as a strategic practice. As explained by P9, \textit{"I wish it was one of the hashtag ones because then I would be like, 'Oh, yeah, I'd probably follow that hashtag.' But like I said, I'm just like literally winging it out here in TikTok, and whatever works, works, and I keep going with that"}. In letting the algorithm take over and disseminate content, creators learned to \textit{have faith} that the algorithm knew the appropriate audience for the content. They identified this form of creation as a low-effort, authentic, and intimate process, also sometimes utilizing spiritual language, as P9 continued:

\begin{quote}
    \textit{I [record] on the spot. Nothing's ever planned. It's \textbf{whatever is aligned at that moment}, and that's how I create them. I just decided I wanted to do that at that moment, got my phone, and I was literally in my PJs. I'm in a t-shirt. That's bare face.}
\end{quote}

We observed that this lack of clarity about the TikTok algorithm strategically led \textit{intuition} to play a role in how creators saw, understood, and performed algorithmic conspirituality. In fact, accepting the limits of their ability to know the algorithm allowed creators to \textit{ignore} the algorithm in their creative process. For instance, P12 hoped for the best, taking somewhat of a "\textit{leap of faith}" as she published her videos- \textit{"It's just, [you] shoot your shot, make the video, put it out there"}. Likewise, P5 explained, \textit{"I don't feel like it's a necessity [to understand the algorithm] because I've been able to kind of do it with just kind of doing free willy-nilly}. P1 affirmed this, stating: \textit{"I've kind of learned [the algorithm] is its own beast. It's like, it changes all the time. Stop trying to figure it out.}" Even when the algorithm's unknowable characteristics frustrated creators, the result was to not focus on it, instead \textit{"let[ting] the universe handle what it wants to handle,' and [they] just really don't look into it" (P9)}. 

%This reflection suggests another way the creator spoke of something beyond herself, bringing in concepts of spirituality while both understanding and presenting algorithmic conspirituality through her intuitive-- non strategy-- strategy.

Instinctive creation and perceived encouragement to create spontaneously intersected with creators' understanding and performance of algorithmic conspirituality. This also fostered acceptance of the algorithm's unpredictability and its serendipitous nature. As P8 noted:\textit{"I think that's why [...]in the content I'm creating, it really helps to not give a crap about it, like to not think about it, to not really give it attention"}. Such discourses were supported by other creators, who were experienced in using TikTok, yet elected not to learn about the algorithm: \textit{"People ask me for my advice and [I] go, 'You know what? I know how to tik. I do not know how to tok.'" (P3)}.

Overall, we note how creators described operating with an explicit absence of strategy. They were cognizant of the importance of performativity, visibility, and other metrics, yet their personal understandings and relational motivations often superseded strategic considerations-- in developing a non-strategy strategy. This approach was understandable, given their perception that strategic practices were never consistently effective due to the algorithm's unknowable nature. This also led to discourses about spiritual and religious motivations, afforded by having \textit{faith} in the algorithm.

\subsection{Creator Consequences: The Impact of Algorithmic Conspirituality}
We observed that creating algorithmic conspirituality content intensified the affective and emotional labor creators performed. Their experiences indicated that the format of algorithmic conspirituality videos invited viewers to feel a sense of intimacy that came with expectations of care and support.

\subsubsection{Affective Labor}
\label{sec:affectivelabor}

Regardless of whether their content explicitly claimed to channel a higher power or offer practical advice, creators highlighted the \textit{"overwhelming"} nature of viewer responses. For instance,  P6 discussed receiving influxes of comments, texts, direct messages, and emails, while P10 noted responding to all the comments would require several hours.

Some further noted that certain viewers developed emotional attachments to them due to the sense of care and compassion their content conveyed. For example, P4 mentioned: \textit{"A lot of people feel immediately like I’m their friend, or they feel comfortable enough to just relate with me."} Similarly, P5 observed that his viewers often formed \textit{“parasocial relationships”} with him, explaining: \textit{“I’m open with people; some of them kind of take it too far, like you give them an inch, they take a mile. I’ve had people do that, and it can be uncomfortable”}.

In some instances, creators reported that viewers not only felt connected to them but also idolized them. P9 recounted how some clients who discovered her through TikTok \textit{"put me [her] on some type of pedestal where they’re like, 'Oh, when I become like you, then I can be happier'"} This idolization could deepen viewers’ attachments, particularly when creators invoked or implied a connection to a higher power-- being motivated by religious or spiritual ideals. For instance, P2 shared:
\begin{quote}
    \textit{[S]ometimes [viewers] take it as if I am like a god or I am a prophet. I’ve gotten people who have messaged me and wanted me to bless things and […] wanted me to just tell them about how things are.}
\end{quote}

As P2 indicates, viewers' parasocial relationship with creators takes on a different valence when deemed religiously or spiritually significant. Understandably, creators reported feeling a tremendous sense of responsibility in their interactions with viewers similar to care professionals \cite{Nourrizetal}. For most creators, an altruistic desire to help others and a sense of connection with viewers drove this responsibility. They viewed their interactions as deeply meaningful, often likening them to relationships with friends, family, or therapists. For example, P12 shared: \textit{I want [viewers] to feel happiness [..] And it’s a big part of also wanting to take care of people’s mental health, right?} This desire to care for people motivated many creators to continue posting and interacting. For example, P4 discussed the many times she considered deleting her account or stopping her posts, but ultimately continued due to the impact she felt she had on her viewers: \textit{"But then I’m like, ‘Dang, I have a lot of people out here that are going through things. And they’ve shown or told me the impact that I’ve had on them. I don’t want to let them down.’"} 

Likewise, creators often found themselves performing labor to manage their own stress and anxiety to support their viewers’ emotional needs. They worried that not posting on any given day could have real consequences. For instance, P2 shared: "\textit{I’ve had so many people reach out to me like, ‘You have no idea how much this is keeping me going.’ So even then, I have this almost anxiety to just continue to put the content out like that.}" In this vein, P3 provided a vivid picture of what he imagined was at stake if he failed to post: 

\begin{quote}
\textit{[I]t’s intense, insanely intense that a message they saw on TikTok, one woman, the very first letter I got, she wrote all of her letters out to her family. That was going to be the night she was going to do it. And so she was scrolling through TikTok and one of my videos popped up and she watched it and she watched every single video that I had as far as the messages go. And she said because of the videos, she didn’t do it. And she’s still around.}
\end{quote}

While P3 described such experiences as instilling him with a sense of purpose, he also lamented: \textit{"it’s heavy. Some of the stuff they say, I don’t even know how to respond to. It’s too much sometimes."} 

The sense of responsibility felt by some creators also followed from the attachments and expectations viewers developed in response to content designed to be motivational or guiding. Some participants' experiences illustrated how content positioned as containing religious or spiritual wisdom heightened the pressure creators felt in response to audience expectations. For example, P6 discussed receiving messages from viewers who were disappointed in the lack of results they saw after following her spiritual advice on "manifestation": 

\begin{quote}
\textit{There was actually one person that I had to tell them like, ‘You need to love yourself more than you love somebody else.’ Because this person was like, ‘If I don’t get this person to text me back, I’m literally going to die.’ And I was like, ‘Oh, my gosh. They’re putting everything, putting all this stuff on me.’ And then I started to feel guilty.}
\end{quote}

Creators' content invited dependence, as viewers valued their predictions, which, when true, validated and deepened that dependence, resembling a form of faith akin to spiritual or religious beliefs. For example, P12, whose niche was tarot card readings, found that some viewers became obsessive in seeking her insight through a client line she established for viewers who bought readings: \textit{"And they would be texting me in the middle of the night-- Oh, my God. --trying to call me in the middle of the night because they were like having a mental breakdown."} Situations like this illustrated the heavy burdens that creators felt, stemming from the connections they developed with their viewers. Notably, this dynamic also mirrors the emotional and affective stress that priests experience in religious settings, where they are expected to bear the emotional weight of others’ struggles \cite{ruiz-prada2021occupational}.

As another example, some of P2’s videos' included specific dates attached to predictions. For example, \textit{"If you’re seeing this video on November 20th, this message is meant for you. Something big is getting ready to enter into your life.”} While he intended such messages to inspire and affirm, he shared viewers often interpreted the content as \textit{divine} predictions, prompting them to seek his advice on matters of great consequence:

\begin{quote}
\textit{[I]t was scary having people message me like, ‘I need you to give me some kind of message to take care of my financial well-being.’ And I’m like, ‘Listen, I can give you some affirmations to help get you in the mindset, but there’s so much more that we got to look at to try to see that happen…’}
\end{quote}

Affirmation as a spiritually motivated emotional response often emerged when individuals seeked guidance in times of crisis, much like turning to traditional religious leaders \cite{10.1145/3686926}. Likewise, this dynamic played out on TikTok, where creators like P2 positioned themselves as sources of insight. Here, the format of P2’s content shaped the emotional responses he received and subsequently had to manage, as it invited viewers to see him as a purveyor of insight from a higher power. In this way, P2's role mirrored that of religious leaders, offering emotional reassurance and spiritual comfort, while navigating the burdens and expectations of his audience.

\subsubsection{Emotional and Digital Labor}

We examined emotional labor as the labor performed by creators to live up to the social expectations of viewers/followers \cite{doi:10.1177/20594364221096498}. While affective labor turned outward to manage others' affective needs, emotional labor turns inward to manage one's own emotions in order to provide good service to "customers." Our participants experienced negative responses from viewers including hate, harassment, and bullying. They also experienced guilt, frustration, and anger as a result of viewers' expectations. Explaining how user expectations about their posts impacted her, P4 shared, \textit{"[I]f I don’t post today and somebody’s going through something, then I could let them down."} 

The perceived imperative to maintain a steady presence on TikTok was taxing, but a necessity signifying creators' own success. Creating motivational, highly personal content based on intuition, without a defined strategy, resulted in creators' opening connections with varied audiences. Consequently, audiences placed substantial value on the creators, depending on and expecting a great deal from them. Not only did this prompt empathy fatigue \cite{Nourrizetal}, but also occasionally invited hostile responses from their audiences. For instance, P3 explained: 

\begin{quote}
    \textit{I want people to think that I’m talking directly to [them]. "Do you feel this message?" I want you to think I’m talking to you. But because sometimes they don’t know how things work, like "How did I get this message? Jesus, did you send me this message? Did I need to hear this today?" And then boom. And let’s say I don’t respond to a DM fast enough to them [...] And it takes me another two weeks to get to them. Oh my God, the things they say in those messages [...] sometimes they get violent. They get verbally assaultive. It’s, yeah, the name-callings.}
\end{quote}

If their absence was felt, reactions sometimes spilled over to the real-world. For example, P3 explained how one time police officers showed up at his door to perform a wellness check after a follower tracked his information down when he did not post for 24 hours. Understandably, P3 described this experience as \textit{"scary"}, but also immediately shifted to emphasize the upsides of cultivating this kind of intimacy with others on TikTok. 

Some creators also shared that they felt pressured to consistently engage with audiences and foster a stronger community. For P13, this pressure grew from comparison to other creators she perceived as \textit{"having strong communities"}. Consequently, she often felt a sense of failure and self-reproach when she perceived herself as falling short of these standards.

\begin{quote}
    \textit{Maybe I'm just being hard on myself, honestly, because I respond to comments.[But] I don't respond to all the comments. I don't have time to respond to all the comments. But I feel like I do a decent job at responding to comments. (P13)}
\end{quote}

Managing emotional labor was often intertwined with performing \textit{digital} labor. Harassment, such as duplicate and fraudulent accounts attempting to scam followers, used creators' identities to deceive others. Although creators were not the \textit{direct} targets, their names and social capital were exploited. Consequently, creators had to manage the fallout, posting clarifications and reaching out to TikTok for assistance. These fake accounts did not just duplicate the name or handle of the creators, hoping for financial gain from misleading followers, but also duplicated content. For example, P10 shared: "\textit{I think I’ve blocked over 30 people trying to be me, and they’re always trying to do tarot card readings and all kinds of stuff.}" 

When creators reached out to platforms for assistance, there was often no action for \textit{"month and months and months"} (P10), and if they did take action, the creators' own accounts were often impacted. P5 added to this, on how platform policies were also not transparent towards her content while allowing fraudulant, mimicry accounts to continue existing, "\textit{TikTok literally flagged my video and took it down, said that I was harassing [...] but they still let the other accounts stay up.}”

Navigating these challenges while also seeking to maintain visibilty and the performativity of producing consequential content was frustrating, exhausting, and difficult to keep up with. As discussed in section \ref{sec:intu}, creators often engaged in a \textit{spontaneous and instinctive} process in maneuvering through the algorithm's minutia. However, the continual unpredictability of the algorithm often left them playing a guessing game that P13 described as \textit{toxic}, sharing: \textit{"It feels like a toxic relationship. Oh, we're doing so good. Oh, this is really bad. Oh, it's actually pretty good, bad. It's back and forth."}

%Such complicated maneuvering, learning, and management while making sense of their content and the algorithm in spiritual and religious manners that elicited emotional responses, added significant emotional and digital labor. 

\begin{table}
    \centering
    %\arraystretch{0.8} % Increase line height for better readability
    \begin{tabular}{|p{0.3\linewidth} | p{0.7\linewidth}|}
        \hline
        \large \textbf{Themes} & \large \textbf{Subthemes} \\
        \hline
        \multirow{3}{*}{\small \parbox{1\linewidth}{\raggedright Displaying \& Performing Algorithmic Conspirituality}} 
        & \vspace{0.2cm} \small A Divine Intervention: Creators attribute their success to divine intervention, believing that a higher power guides their actions and content. They feel that their creativity and messages are inspired by spiritual and religious forces, whether through a direct connection to God or other transcendent experiences. This belief drives them to produce content with a sense of purpose, trusting that it will reach the right audience, often relying also on the role of the algorithm to deliver their messages. \vspace{0.2cm}\\
        \cline{2-2}
        & \vspace{0.2cm} \small Personalized Engagement: Personalized Engagement refers to creators using a "you-centric" approach in their algorithmic conspirituality content to create a personal connection with viewers, enhancing audience engagement by making the message feel specifically tailored to individual needs and increasing the likelihood of interaction. \vspace{0.2cm}\\
        \cline{2-2}
        & \vspace{0.2cm} \small Intuitive Spontaneity: Intuitive Spontaneity refers to a creative process where creators embrace a lack of predetermined planning, allowing inspiration and intuition to guide their content. This "non-strategy strategy" often relies on trusting the algorithm’s unpredictability, creating content spontaneously and authentically, without overthinking or focusing on strategic planning, leading to a sense of serendipity and faith in the process. \vspace{0.2cm}\\
        \hline
        \multirow{2}{*}{\small \parbox{1\linewidth}{\raggedright Creator Consequences}} 
        & \vspace{0.2cm} \small Affective labor emerged from algorithmic conspirituality as a result of the format and genre of the content created. Creators were required to provide emotional and psychological support to their audiences. This involved managing parasocial relationships, addressing viewers' emotional needs, and acting as a source of guidance. Creators often felt a heavy obligation to their viewers, similar to that of care professionals or religious figures, especially when seeking emotional validation or spiritual reassurance. This dynamic led to both positive and overwhelming forms of emotional labor. \vspace{0.2cm} \\
        \cline{2-2}
        & \vspace{0.2cm} \small Emotional and digital labor arose as creators managed the emotional expectations of their audiences, including dealing with harassment and guilt, while also handling the fallout from fraudulent accounts and platform inconsistencies, all while navigating an unpredictable algorithm.
        \vspace{0.2cm}\\
        \hline
    \end{tabular}
    \caption{A Summary of the Key Findings}
    \label{tab:findingsum}
\end{table}

\section{Discussion}

Our findings reveal creators' diverse interpretations of algorithmic conspirituality content and the motivations behind its creation. Some creators viewed their content as spiritually or religiously driven, attributing viral success to divine intervention and framing their work through personal beliefs, suggesting that a higher power guided their content to the right audience. Others saw their algorithmic conspirituality content as a deliberate stylistic strategy. For these creators, phrases like "If you're seeing this, it's meant for you" were employed to engage viewers with personalized language, cultivating a sense of intimacy and connection. Moreover, for many creators, an \textit{intuitive spontaneity} alternately represented an ambivalent practice forming a \textit{non-strategy strategy}. Here, creating content was based on \textit{faith} in the (algorithmically-mediated) powers that be. Regardless of their understanding of and motivations for creating algorithmic conspirituality content, many creators reported overwhelming, invasive viewer interactions due to the expectations the genre of content invited, particularly when the content was of religious and/or spiritual nature. Creators' shared that viewers frequently took their messages to heart and idolized them, as more than human, including sometimes as messianic figures. In the following sections, we situate our contributions within CSCW, focusing on content creation, technospiritual discourse—specifically, the integration of religious and spiritual concepts in HCI—and the associated labor.

\subsection{Religion, Spirituality, Faith and Content Creation}

%\subsection{Algorithmic Conspirituality: Strengthening Faith and Connections}

Our work highlights the often unexamined traces of religion and spirituality in everyday technology use, helping to expand the underdeveloped domain of HCI research on religion and spirituality \cite{10.1145/3613905.3651058,10.1145/2468356.2468754}. We uniquely demonstrate how spiritual and religious beliefs may become organically, perhaps unexpectedly, imbricated with social media affordances, and user culture. This contradicts a popular view of technology as incompatible with spirituality and religion \cite{10.1145/3613905.3651058}, instead drawing attention to the possibilities of their entanglement. In fact, we assert that it is the nature of algorithmic conspirituality based content, that makes it appealing to religious, spiritual, and faith-based followers. The characteristic terminology for this content alludes to being "meant for" users, implying a certain outerworldly belief in directing this content to them \cite{doi:10.1177/14614448231217425}. As various HCI scholars have documented, social media creators make sense of algorithms in diverse ways (See for eg., \cite{10.1145/3610169, DeVito_algotrap}). Adding to this work, our findings flag religion, spirituality and other types of existing alternative epistemological beliefs as relevant to this sensemaking.

%We find that preexisting beliefs often contribute to the creation of algorithmic conspiratorial content, particularly to reinforce and disseminate these convictions.

Characterized by its perceived serendipity \cite{doi:10.1177/20563051221077025}, TikTok is ripe for the propagation of content incorporating ideas of religion and spirituality \cite{partridge2021when}, consistent with prior work highlighting that creators' attraction to unpredictability drives their preference for the platform \cite{10.1145/3555578}. We extend CSCW research on creators' sensemaking and content creation practices, as well as the role of religious and spiritual beliefs in technology use, by demonstrating how creators' spiritual and/or religious beliefs, combined with their perceptions of the algorithm's serendipitous nature, inspire the creation of content within the algorithmic conspirituality genre. This happens as \textbf{serendipitous views of the algorithm reinforce beliefs}, leading to \textit{intuitive spontaneity} as a \textit{non-strategy-strategy}. This intuitive spontaneity acts as an important characteristic of the content such creators create- consequently leading to the relationships, platform perceptions and support that they both provide and expect from the content creation and sharing process.

 %Past work has theorized \textit{reflexive ambivalence}, wherein users perceive the TikTok algorithm, particularly in how it presents algorithmic conspirituality content, as both imbued with a sense of mystical agency and in tension with users' awareness of the data-driven processes behind recommendation algorithms \cite{doi:10.1177/13548565241258949}. 

Consequently, we extend research on creators' audience engagement strategies \cite{ma_multi-platform_2023} by exploring how creators develop ideas spontaneously, without relying on specific strategies to navigate TikTok's algorithm. It helps that TikTok's seamless flow of content obscures the underlying algorithmic processes, inviting creators to explicitly characterize their content as appearing in front of users as a result of a higher power. Whether creators mean this in earnest or in a tongue-in-cheek manner, TikTok's serendipitous "right time, right place" algorithmic recommendations \cite{Zhang2021TikTok, 10.1145/3613904.3642297, 10.1145/3555601} invite a sense of connectedness between viewers and creators. Here, as noted in \cite{10.1145/3613905.3650743}, when situating algorithmic systems within a more than human plane, their use can prompt introspection and a sense of \textit{"emotional solidarity"} with others who become present through mutual interaction with the system. Indeed, the intense and sometimes intrusive responses described in section \ref{sec:affectivelabor} exemplify the deepening of algorithmic conspirituality content's connective affordance \cite{doi:10.1177/14614448231217425}, as premised on a notion that everything is connected.  

% For both viewers and creators, the experience of being connected in this way creates space to fortify religious, spiritual and other more than rational beliefs about the platform's algorithm and, by centering faith, may minimize demands for greater transparency or explanation. 

\subsection{\textbf{This Message is Meant for you}: Characteristics of Algorithmic Conspirituality Content}

Our work also builds on research contributing to the creator economy's focus on creating personalized experiences \cite{10.1145/3555601, 10.1145/3544548.3581386} by explicating how content, influenced by spiritual and religious beliefs also contributes to its commercialization. In particular, we explain how the design of content, shaped by creators' beliefs and their interpretation of the TikTok algorithm, draws on communication practices that aim to be personalized, yet are also designed for the masses. Prior research has highlighted tensions between authenticity and visibility, particularly due to algorithmic opacity \cite{10.1145/3613904.3642173, doi:10.1177/1461444818815684}. We demonstrate how algorithmic conspirituality content implicitly mediates these tensions, despite a lack of explicit concern about them.

Among our interviewees, we found that creators' content, while not claimed for specific audiences, phrased words that explicitly only spoke to a specific individual or group that they believed would receive the content-- be it through algorithmic or outer-worldly mediation. This was exemplified in \ref{sec:peson}, where P14 imagined his audience as all who needed to \textit{hear the message}, and P10 and others focused on reaching audiences by using "you" to make it more personalized and engaging. From this, it is evident that creators perceived the phrase "this message is meant for you" and similar phrases as powerful methods of influencing the masses through the personalized structure of the content. Many creators intended this phrase literally, rooted in their belief in a higher power, while others adopted it as a persuasive hook to engage viewers. Likewise, the message exhibits \textit{interpretive flexibility}-- depending on viewers' distinct experiences, perspectives, and expectations, it influences how one interprets the message and makes sense of it, according to what they see as right (see \cite{doi:10.1177/13548565241258949}). However, the algorithmic conspirituality genre does embed an open-ended spirituality,  \textit{accommodating major religious traditions, new age beliefs, or agnostic viewpoints}-- as not being mutually exclusive to any particular beliefs. This places creators in a distinctive position when it comes to persuasive appeals \cite{10.1145/2209310.2209312}, as much like the \textit{personalized economy}- the message is produced for mass consumption but also serves as \textit{customized products for individualistic consumers} \cite{russell1993personalized}. This reflects the ambiguous spirituality of algorithmic conspirituality, aligning with the contemporary shift from religion to spirituality, which emphasizes a more personal religious experience \cite{einstein2008brands}. In this context, belief is influenced, shaped, or amplified by algorithms, often in complex, contradictory, and unclear ways-- as seen through the varied experiences and performances of algorithmic conspirituality. 

In a way, we see changes in how spirituality and religion are used as forms of support, as creators generate "you-centric" content, and the algorithm's personalization factor strengthen this. Consequently, the TikTok algorithm, becomes an artifact of \textit{faith-ing} \cite{hoefer2022faith}- supporting goals of faith informatics such as making connections and maintaining and sharing faith. These changes reflect traits of a personalized economy \cite{russell1993personalized} but also facets of the creator economy, where creators are interested in being authentic, diverse, and novel \cite{10.1145/3613904.3642173, duffy2021value} while also needing to produce content constantly \cite{doi:10.1177/2056305120944624}, leading to the mass production of personalizable content within the spiritual realm. The mass production of personalized spiritual and religious content extends to leaders of institutionalized religions, who engage audiences on algorithmically driven platforms while navigating tensions between social media norms and traditional communication practices \cite{campbell2020digital}.

\subsection{Affective Labor: Content Driven Responsibility}

Our findings underscore creators' affective labor within the constraints of an algorithm-driven creator economy, extending prior research on digital content production (e.g., \cite{cummings2019but, doi:10.1177/20594364221096498}). We find that the genre of content created a pronounced sense of responsibility for creators, resulting in a ubiquitous obligation regardless of their initial intentions.

Participants' reflections drew parallels to \textit{empathy fatigue}, or \textit{"a state of emotional, mental, physical, and occupational exhaustion"} \cite{stebnicki_empathy_2007} prompted by efforts to understand and connect with others' experiences, experienced by spiritual care, medical, and counseling professionals, among others\cite{Nourrizetal, stebnicki_empathy_2007}. Some participants' attempts to enact boundaries with their viewers demonstrates a need for tools and resources for striking the right balance between positive and negative impacts of creators' empathy\cite{Nourrizetal}. We also extend prior work that has shown online spiritual support can be beneficial in various contexts, including but not limited to supporting individuals' mental, physical, and emotional health \cite{10.1145/3449117, keating2013spirituality}, as well as advising people in everyday decision-making \cite{wyrostkiewicz2022catholic}. Our findings explored interviewees positioning themselves as providing such support, viewing themselves as influential figures and guiding lights in their audiences' lives through their content. Once they carve out such a niche—often through a viral video—they take on accountability for their content and its impact. Affective labor hence manifests through the consequent \textit{affective relationship} with their audiences. As P4 explained, \textit{I have a lot of people out here that are going through things. And they’ve shown or told me the impact that I’ve had on them. I don’t want to let them down}- in that the very idea of her content- produces \textit{relationships and emotional responses} (\cite{articleosaka}; p. 284)- which are products of affective labor \cite{doi:10.1177/1354856517736983, hardt1999affective}. Consequently, ambiguous spirituality through individualized messaging adds intimacy \cite{doi:10.1177/1354856517736983} to this affective labor, producing a strengthened reaction from viewers and deepening the emotional expectations from creators.

Our analysis reveals that affective labor in this genre arises from individualized messaging, particularly in its spiritual or religious dimensions. This highlights the complexity of content creation, requiring creators to engage deeply with followers while managing personal boundaries and identities. The expectations of social intimacy, guidance in everyday decision making, and opportunities for fellowship is similar to what has been seen in traditional congregational use of social media for ministry work \cite{campbell2019contextualizing, osti_10376103}-- extending to Kong's \cite{kong2003religion} work who theorized that the blending of mass and personal communication would facilitate these types of dependencies in religious spaces. However, creators reported their followers developed strikingly high levels of interpersonal relational expectations. As creators reported these experiences were more pronounced on TikTok, our findings also align with work that indicates the nature of more interpersonally-focused relationships between spiritual leaders and followers is dependent on context and platform \cite{campbell2019contextualizing}. These findings suggest a need for further exploration of the the experiences of traditional religious leaders who have cultivated social media followings who may have unique perspectives on differences between their traditional interpersonal relationships with congregants and parasocial relationships with their social media community.

\subsection{Implications for Platform Design}

Prior work has provided design suggestions to support religious and/or spiritual contexts, encouraging ideas such as openness, fluidity, and uncertainty in the design process \cite{spirited_collective}. While these concepts inform platform design for religious and spiritual needs, our work examines content creators' challenges in algorithmically mediated systems. Increased algorithmic accuracy reinforces users' perceived dependence on content timing and fosters expectations of creator attention and care. Scholars have also highlighted spirituality as a key motivation for engagement on platforms \cite{10.1145/3449117}. Addressing how to mitigate the labor of content creators in this space is crucial for sustaining their work and supporting user needs.

Content moderation tools can ofcourse help reduce bullying and harassment faced by creators \cite{10.1145/3491102.3501879, 10.1145/3613904.3641949}, but our findings suggest that additional tools are needed to support creators' emotional and affective labor. Creators do not want to eliminate viewer interactions, but seek more effective ways to manage them without being constantly available.

Platform design for mediating affective labor should prioritize creating tools and resources that \textit{"respect the emotional limitations"} \cite{Nourrizetal} of creators, helping them manage their own emotional efforts while simultaneously supporting their ability to care for their audiences. The goal here is sustainable labor, which necessarily must include opportunities to disengage and recuperate \cite{Nourrizetal}. For instance, TikTok could incorporate a tool that enables creators to acknowledge multiple messages simultaneously, thereby reducing the pressure to respond individually to each one. This could include automated acknowledgment tools, such as a button to acknowledge or auto-respond, conveying attentiveness to the audience and affirming that their messages are valued. It would also continue to maintain communities and relationships with audiences, which would be mutually beneficial for both the platform and creators \cite{10.1145/3579477}. In the same vein, the video reply feature, which allows creators to respond to individual comments in a video, could be enhanced to select comments expressing similar sentiments and tag all commenters in the response video. Alleviating the \textit{anxiety} and \textit{overwhelming responsibility} experienced by creators not only fosters their creative processes by minimizing the burden of affective and emotional labor but also enhances their relationships with users who seek meaningful connections with them \cite{10.1145/1940761.1940767}. 

Notably, these recommended designs represent less effortful means of interacting with audiences, which could attenuate perceived presence and connection \cite{zhangetal}. While it is important to give creators tools that grant a reprieve from the demands of affective and emotional labor, creators will need to understand the trade-off that comes with employing them and implement them not as default responses but situationally, as appropriate. Moreover, platforms should take care to recognize the differences between reducing creators' \emph{personal efforts} related to maintaining relationships and \emph{procedural efforts} related to navigating and exploiting platform features \cite{zhangetal}. As \cite{zhangetal} show, reducing personal effort can be achieved by building meaningfulness via design elements that make communication akin to a gifting experience: personalization, cuing the notion of wrapping and unwrapping, and making effort visible to receivers. Meanwhile, reducing procedural effort can be achieved through designs that promote playfulness in communicative tools, intuitiveness, and asynchronicity \cite{zhangetal}.

Our findings demonstrate creators simultaneously seeking to cultivate a favorable relationship with the algorithm and audiences. While serendipity as a non-strategy-strategy remains vital in content creation, and supporting other work, and sharing the exact workings of algorithms remains difficult, reducing the labor required for creators to engage with their audience is of value. A system that enables viewers to provide nuanced feedback—similar to Facebook’s reactions but with options like \textit{hopeful}, \textit{cared for}, or \textit{anxious}-- could streamline interactions, allowing creators to better understand audience sentiment without being overwhelmed by individual user needs. By minimizing the effort required to interpret audience responses, such a system would support creators in shaping content that remains spontaneous, responsive to audiences' need and an effective source of support. Ideally, creators would be able to customize viewer response options that provide them with information relevant to their unique goals and communities. However, the flexibility that customizable responses afford must be balanced with increased cognitive demand for viewers who must determine the most appropriate reaction graphic to select as well as the ambiguity of what each option represents \cite{smithetal-thoughts}.

\section{Limitations, Conclusion and Future Research}

%we contacted large number- not representative- we reached saturation- specific kind of phenomenon- no option to look at generic

Our research has limitations. As with all qualitative studies, our findings reflect the research team's interpretation and are not intended for generalization \cite{Braun2019, Braun2019a}. Data collection ceased upon reaching saturation through iterative analysis and discussion \cite{Braun2019, Braun2019a}, though further interviews could yield additional insights, especially as we are limited by geographic diversity, with all interviewees based in the US, which may influence their beliefs and experiences. As Cotter et al. noted, spiritual sensemaking in algorithmic conspirituality may be shaped by the American tradition of religious liberalism \cite{doi:10.1177/13548565241258949}, intersecting with political liberalism \cite{schmidt2012restless}. We encourage future research to explore diverse cultural perspectives. Lastly, our interviews focused on TikTok; further investigation into platform variations and algorithmic conspirituality is needed. We also point to the predominantly spiritual and religious content observed in searches for algorithmic conspirituality videos and call for future work to further center such content and experiences in platform studies.

In conclusion, our research highlights the spiritual, relational, and strategic motivations behind algorithmic conspirituality content. Creators relied on intuitive spontaneity, often embedding religious and spiritual ideas about algorithms and trusting them to reach the right audience. While creators described this approach as a means of connecting with and supporting viewers, they reported it often led to overwhelming responses. Our study advances HCI and CSCW research by showing how creators' beliefs, motivations and TikTok's algorithm shape digital community dynamics and perceptions of algorithmic conspirituality. We also reveal how individualized "you"-centric communication on TikTok fosters support, underscore the relational nature of persuasive content and bring out spiritual ambiguity as a consequence of such content. Finally, we propose design tools to manage affective labor in spiritual and religious content creation on TikTok, providing a foundation for other researchers to also support similar content within platformized systems.

\bibliographystyle{ACM-Reference-Format}
\bibliography{references.bib}

%\appendix
%\input{appendix}

%\appendix

%\section{Appendix header}

\end{document}